\documentclass[prd,onecolumn,superscriptaddress]{revtex4}

\usepackage{graphicx}

\usepackage{hyperref}
\usepackage[english]{babel}
\usepackage{amsmath}

\newcommand{\mrm}[1]{\mbox{\rm #1}}
\newcommand{\be}{\begin{equation}}
\newcommand{\ee}{\end{equation}}
\newcommand{\br}{\begin{eqnarray}}
\newcommand{\bea}{\begin{eqnarray}}
\newcommand{\eea}{\end{eqnarray}}
\newcommand{\er}{\end{eqnarray}}
\newcommand{\ba}{\begin{array}}
\newcommand{\ea}{\end{array}}
\newcommand{\bi}{\begin{itemize}}
\newcommand{\ei}{\end{itemize}}
\newcommand{\bn}{\begin{enumerate}}
\newcommand{\en}{\end{enumerate}}
\newcommand{\bc}{\begin{center}}
\newcommand{\ec}{\end{center}}

\newcommand{\nn}{\nonumber\\}

\newcommand{\Eq}[1]{Eq.~(\ref{#1})}
\newcommand{\rfn}[1]{(\ref{#1})}

\newcommand{\gev}{\mbox{GeV}}

\newcommand{\gsim}{\lower.7ex\hbox{$\;\stackrel{\textstyle>}{\sim}\;$}}
\newcommand{\lsim}{\lower.7ex\hbox{$\;\stackrel{\textstyle<}{\sim}\;$}}

\newcommand{\hc}[1]{#1^{\dagger}} 
\newcommand{\abs}[1]{|#1|}

\begin{document}


\title{\bf 
 Dark Matter as the signal of Grand Unification
}

\author{ Mario Kadastik}
\author{ Kristjan Kannike}
\author{ Martti Raidal}

\affiliation{National Institute of Chemical Physics and Biophysics, Ravala 10, Tallinn 10143, Estonia}

\pacs{}

\begin{abstract}
We argue that the existence of Dark Matter (DM) is a possible consequence of GUT symmetry breaking. 
In GUTs like $SO(10),$ discrete $Z_2$ matter parity  $(-1)^{3(B-L)}$ survives despite of broken $B-L,$ 
and group theory {\it uniquely} determines that the only possible $Z_2$-odd matter multiplets belong to representation $\bf 16.$ 
 We construct the minimal non-SUSY $SO(10)$ model containing one scalar $\bf 16$ for DM and study its predictions below $M_{G}.$
We find that EWSB occurs radiatively due to DM couplings to the SM Higgs boson.
For thermal relic DM the mass range $M_{\mathrm{DM}}\sim {\cal O} (0.1-1)$~TeV is {\it predicted} by model perturbativity up to $M_{G}.$
For $M_{\mathrm{DM}}\sim {\cal O}(1)$~TeV to explain the observed cosmic ray anomalies with DM decays,
there exists a {\it lower bound} on the spin-independent direct detection 
cross section within the reach of planned experiments.
\end{abstract}


\maketitle

\section{Introduction}
 The existence of Dark Matter (DM) of the Universe is now established without doubt~\cite{Komatsu:2008hk}.
However, the fundamental physics behind it is unknown at present.
In the most popular new physics scenario containing DM -- supersymmetry -- discrete $R$-parity is imposed
by hand to prevent phenomenological disasters such as fast proton decay~\cite{Farrar:1978xj}.
Similarly, in dedicated DM extensions of the standard model (SM) with new singlet~\cite{rs}, 
doublet~\cite{id} or higher multiplet scalars~\cite{Hambye:2009pw}, {\it ad hoc} $Z_2$ symmetry
must be added to ensure the stability of DM.
These phenomenological models cannot answer the two most fundamental questions related to DM: 
$(i)$ why this particular multiplet or particle constitutes the DM of the Universe?; 
$(ii)$ what is the origin of the imposed $Z_2$ symmetry?
Therefore the  underlying physics principles related to the existence of DM  remain obscured.

In this work we argue that the existence of DM of the Universe can be a consequence of
Grand Unification (GUT). The GUT framework not only explains the origin of DM but also  
determines the type of the DM particle and constrains its properties. 
In this scenario the existence of DM, non-zero neutrino masses via seesaw~\cite{seesaw} 
and baryon asymmetry of the Universe via leptogenesis~\cite{Fukugita:1986hr} 
all point to the same GUT framework.

We show that the $Z_2$ symmetry needed for DM stability could be a discrete remnant of GUT symmetry group,
such as $SO(10)$~\cite{Fritzsch:1974nn}
 that we choose to work with in the following. When breaking $SO(10)$ 
down to the SM gauge group $SU(2)_L\times U(1)_Y,$ the $SO(10)$ embedded $U(1)_X,$ 
where $X$ is orthogonal to the SM hypercharge $Y,$ leaves unbroken $Z_2$~\cite{Krauss:1988zc,Martin:1992mq}
\bea
 P_X=P_M=(-1)^{3(B-L)},
 \label{P}
\eea
which is the well known matter parity $P_M.$ Due to its gauge origin $P_M$ is a symmetry of
{any} SM extension including non-SUSY ones. In the latter case group theory predicts
{\it uniquely,} without any detailed model building, that the only possible $Z_2$-{odd} multiplet under
\Eq{P} is the ${\bf 16}$ of $SO(10)$~\cite{Kadastik:2009dj}. As inclusion of the fourth fermion generation ${\bf 16}_4$ to the SM
is not supported by experimental data, the non-SUSY $SO(10)$ GUT {\it predicts}
that the DM is a mixture of $SU(2)_L\times U(1)_Y$ 
 $P_M$-odd complex scalar singlet $S$ and neutral component of doublet $H_2$
belonging to a new scalar ${\bf 16}$ of $SO(10).$
 Thus the DM of the Universe corresponds to the 
scalar analogues of the fermionic neutral matter fields, the right-handed neutrino $N_R$ 
and the left-handed neutrino $\nu_L,$ respectively.  Preserving $P_M$ requires 
$SO(10)$ breaking by an order parameter carrying even charge of 
$B-L$~\cite{Krauss:1988zc,Martin:1992mq}. Therefore $SO(10)$ breaking also generates 
heavy Majorana masses  which induce the seesaw mechanism  as well as 
 leptogenesis.

To test the proposed DM scenario we study the scalar potential of a minimal $SO(10)$ GUT model 
containing one scalar ${\bf 16}$ for the DM and one scalar ${\bf 10}$ for the SM Higgs doublet. 
We derive \cite{RGE} renormalization group equations (RGEs) for scalar mass parameters $\mu_i^2$ and interaction couplings
$\lambda_i$ below the GUT breaking scale and study the vacuum stability and perturbativity conditions for
those parameters. We find that the SM Higgs mass parameter $\mu_1^2$ runs negative due to 
the presence of DM couplings with Higgs boson and triggers {\it radiative} electroweak symmetry breaking (EWSB) as
in SUSY models \cite{Ibanez:1982fr}. Perturbativity up to $M_G = 2 \times 10^{16}~\gev$ restricts all scalar self couplings to be $\lambda_i < 1$ at $M_Z$ 
predicting a restricted mass window 70~GeV$\lsim M_{\mathrm{DM}} \lsim $2~TeV for thermal relic DM mass. 
The operator
$ m/(\Lambda_N M_P) LLH_1H_2$ induces 3-body decays 
$DM\to l^-\nu W^+$ which may explain the recently observed cosmic ray anomalies.
For the DM mass preferred by this solution, $M_{\mathrm{DM}}\sim {\cal O}(1)$~TeV, our framework
predicts {\it a lower bound} on the spin independent direct 
cross section of DM with nuclei, which is within the reach of sensitivity of proposed experiments.

\section{DM, leptogenesis and seesaw mechanism} 
The $SO(10)$ gauge group contains two orthogonal $U(1)$ charges, which can be chosen to be
the SM hypercharge $Y$ remaining unbroken after $SO(10)$ breaking at $M_G$,
and broken $X=3 (B-L)+4 T_{3R},$ where $T_{3R}$ is the third component of $SU(2)_R\in SO(10)$ isospin.
If $SO(10)$ is broken by fields with even $X$ charge, the discrete subgroup $Z_2$ of
$U(1)_X$ remains unbroken~\cite{Krauss:1988zc}.
As $4T_{3R}$ is always even, the surviving 
$P_X$ parity is nothing but the matter parity, \Eq{P}. Due to the $SO(10)$ breaking, $B-L$ is broken at GUT scale generating
large Majorana masses for right-handed neutrinos $N_{R_i}$ which
suppress the light neutrino masses via the seesaw mechanism.
The $N_{R_i}$ decays in early Universe induce the baryon asymmetry via leptogenesis. 
Thus, in our model, the existence of DM due to the
matter parity \Eq{P}, the existence of baryon asymmetry and the existence of seesaw suppressed masses of light neutrinos 
have the same GUT origin. 
To our knowledge, $U(1)_X,$ $X=5(B-L)-2Y$~ \cite{Wilczek:1979et}, has been used to forbid proton decay 
operators in GUTs, explicit examples of gauged $U(1)_{B-L}$ SUSY seesaw models generating 
$R$-parity have been presented in~\cite{Martin:1996kn} and low energy non-SUSY SM extension with extra $U(1)'$
gauge symmetry generating $Z_2$ is presented in~\cite{Kubo:2006rm}, but the connection between non-supersymmetric 
DM and $P_X$ was first proposed in~\cite{Kadastik:2009dj}.

 The $P_X$ parity of $SO(10)$ matter multiplets is {\it uniquely} determined by group theory. Therefore
the proposed $SO(10)$ GUT scenario leaves {\it no choice} as to what are the DM particle multiplets.
Under the group theoretic decomposition $SU(5) \times U(1)_X$
the ${\bf 16}$ representation of $SO(10)$ reads
${\bf 16} ={\bf 1}^{16}(5)+{\bf \bar 5}^{16} (-3) + {\bf 10}^{16} (1),$ where the $X$ charges
of the component multiplets are given in the brackets. While the $X$ charges are different, all the 
fields in ${\bf 16}$ of $SO(10)$ are {\it odd} under the conserved $Z_2$ parity \Eq{P}.
Interestingly, fields in ${\bf 16}$ provide the only $P_X$ odd particles because 
all other fields coming from small $SO(10)$ representations, 
${\bf 10} ,$ ${\bf 45} ,$ ${\bf 54} ,$ ${\bf 120} $ and ${\bf 126} ,$ are even under $P_X.$
Thus the SM fermions belonging to ${\bf 16} _i,$ $i=1,2,3,$ are all $P_M$-odd while the SM Higgs boson doublet is $P_M$-even
because  ${\bf 10} ={\bf 5}^{10}(-2)+{\bf \bar 5}^{10} (2). $ 

Although $B-L$ is broken in Nature by the heavy neutrino Majorana masses, discrete
$(-1)^{3(B-L)}$ is respected by the interactions of {\it all} matter fields.
Therefore, without any model building, general GUT group theoretic argument 
implies that the non-supersymmetric DM must belong to ${\bf 16}$ of $SO(10).$
Adding a new fermionic $\bf 16$ is equivalent to adding a new generation, which, 
due to mixing with lighter generations, cannot give DM.
The only possibility is the new scalar $\bf 16$ of $SO(10),$
which contains two DM candidates, the complex singlet $S$
and the neutral component of the doublet $H_2.$

\section{Minimal SO(10) GUT induced DM model}
The $SO(10)$ symmetric scalar potential of one $\bf 16$ and one $\bf 10$, \bea
V&=& 
\mu_{1}^{2} \;{\bf 10\:} {\bf 10} +  \lambda_{1} ( {\bf 10\;} {\bf 10})^2 + \mu_{2}^{2}\; \overline {\bf 16} \,{\bf 16} + 
 \lambda_{2} ( \overline{\bf 16}\, {\bf 16})^2 \nn 
&+&
 \lambda_{3}  ({\bf 10\;} {\bf 10}) (\overline{\bf 16}\, {\bf 16} )+ 
 \lambda_{4} ( {\bf 16\;} {\bf 10}) (\overline{\bf 16}\, {\bf 10} ) \nn 
 &+&
  \lambda'_{S}  \left[  {\bf 16}^4 + \mathrm{h.c.} \right] + \frac{\mu'_{S H}}{2}  \left[ {\bf 16\;} {\bf 10\;} {\bf 16} + \mathrm{h.c.} \right],
 \label{VGUT}
\eea
provides the minimal example of 
GUT DM model. Here we have taken all parameters to be real for simplicity.
We assume $SO(10)$ to break at $M_G$ down to $SU(2)_L\times U(1)_Y\times P_M$ in
such a way that only one SM Higgs boson doublet $H_1\in {\bf 10}$ and the DM candidate
complex singlet $S\in {\bf 16}$ and the inert doublet $H_2\in {\bf 16}$ are light, with all
other particle masses of order $M_G.$ 
The $SO(10)$ symmetry breaking may occur in one or in several steps through intermediate
symmetries such as $SU(5)\times U(1)_X.$ We assume those steps to occur 
close to the GUT scale. 
Thus, between $M_G$ and the EWSB scale $M_Z$,
the DM is described by the $H_{1} \to H_{1}$, $S \to -S,$ $ H_{2} \to -H_{2}$ 
invariant scalar potential
\begin{equation}
\begin{split}
V &= \mu_{1}^{2} \hc{H_{1}} H_{1} + \lambda_{1} (\hc{H_{1}} H_{1})^{2} 
+ \mu_{2}^{2} \hc{H_{2}} H_{2} + \lambda_{2} (\hc{H_{2}} H_{2})^{2} \\
&+ \mu_{S}^{2} \hc{S} S + \frac{\mu_{S}^{\prime 2}}{2} \left[ S^{2} + (\hc{S})^{2} \right] + \lambda_{S} (\hc{S} S)^{2} 
 \label{V}\\
& + \frac{ \lambda'_{S} }{2} \left[ S^{4} + (\hc{S})^{4} \right] 
 + \frac{ \lambda''_{S} }{2} (\hc{S} S) \left[ S^{2} + (\hc{S})^{2} \right] \\
&+ \lambda_{S1}( \hc{S} S) (\hc{H_{1}} H_{1}) + \lambda_{S2} (\hc{S} S) (\hc{H_{2}} H_{2}) \\
&+ \frac{ \lambda'_{S1} }{2} (\hc{H_{1}} H_{1}) \left[ S^{2} + (\hc{S})^{2} \right]
+ \frac{ \lambda'_{S2} }{2} (\hc{H_{2}} H_{2}) \left[ S^{2} + (\hc{S})^{2} \right]
 \\
&+ \lambda_{3} (\hc{H_{1}} H_{1}) (\hc{H_{2}} H_{2}) + \lambda_{4} (\hc{H_{1}} H_{2}) (\hc{H_{2}} H_{1}) \\
&+ \frac{\lambda_{5}}{2} \left[(\hc{H_{1}} H_{2})^{2} + (\hc{H_{2}} H_{1})^{2} \right] \\
&+ \frac{\mu_{S H}}{2} \left[\hc{S} \hc{H_{1}} H_{2} + \mathrm{h.c.} \right]
+ \frac{\mu'_{S H}}{2} \left[S \hc{H_{1}} H_{2} + \mathrm{h.c.} \right], 
\end{split}
\end{equation}
together with the GUT scale boundary conditions
\bea
&\mu_1^2(M_{G})>0,\; \mu_2^2(M_{G})=\mu_S^2(M_{G}) >0, & 
\label{bc1}\\
&\lambda_2(M_{G})=\lambda_S(M_{G})=\lambda_{S2}(M_{G}),\; \lambda_3(M_{G})=\lambda_{S1}(M_{G}), \nonumber&
\eea
and
\bea
&{\mu}_S^{\prime 2}, \; {\mu}_{SH}^{ 2} \lsim {\cal O}\left( \frac{M_G}{M_P}\right)^n \mu^2_{1,2},& 
\label{bc2}\\\
& \lambda_{5} ,\; \lambda'_{S1} ,\; \lambda'_{S2} ,\; \lambda''_{S} \lsim {\cal O}\left( \frac{M_G}{M_P}\right)^n \lambda_{1,2,3,4}.&
\nonumber
\eea
We require $\mu^2_i(M_G)>0$ in order not to break the SM gauge symmetry spontaneously at GUT scale.
While the parameters in \Eq{bc1} are allowed by
$SO(10),$ the ones in \Eq{bc2} can be generated only after $SO(10)$ breaking by  operators
suppressed by $n$ power of Planck scale $M_P.$
If all parameters in \Eq{bc2} vanished identically, Peccei-Quinn (PQ) symmetry would imply degenerate
real and imaginary components of DM. However, direct search for inelastic DM requires the mass splitting to exceed ${\cal O}(100)$~keV. 
Smallness of $\lambda_5,$ as given by \Eq{bc2}, allows one to interpret the annual modulation observed by DAMA experiment
with inelastic scattering of DM in the Inert Doublet Model~\cite{Arina:2009um}. 
In our model there are more possibilities to obtain small mass splitting between real and imaginary components of 
DM candidates, {\it cf.} \Eq{V}.
In the following we assume the PQ 
symmetry to be broken {\it softly} by $0< |\mu_S^{\prime 2}|\ll |\mu_1^2|.$

\section{Vacuum stability constraints}
In the SM the requirements of vacuum stability and scalar potential perturbativity up to $M_G$ 
put the lower and upper bounds on the Higgs boson mass $127~\gev<M_H<170~\gev$, respectively (see \cite{stability1} and references therein).
In our model the vacuum stability requires
\begin{equation}
\begin{aligned}
\lambda_{1} &> 0, 
\qquad &\lambda_{3} &> -2 \sqrt{\lambda_{1} \lambda_{2}}, \\
\lambda_{2} &> 0,  
\qquad &\lambda_{3} + \lambda_{4} - \abs{\lambda_{5}} 
&> -2 \sqrt{\lambda_{1} \lambda_{2}}, \\
\lambda_{S} + \lambda'_{S} &> \abs{\lambda''_{S}}, 
\qquad & 8 (\lambda_{S} - \lambda'_{S}) \lambda'_{S} 
&> \lambda_{S}^{\prime\prime 2}, \\
4 \lambda_{1} (\lambda_{S} + \lambda'_{S} + \lambda''_{S}) 
&> (\lambda_{S1} + \lambda'_{S1})^{2},
\qquad & 4 \lambda_{2} (\lambda_{S} + \lambda'_{S} + \lambda''_{S}) 
  &> (\lambda_{S2} + \lambda'_{S2})^{2},\\
4 \lambda_{1} (\lambda_{S} + \lambda'_{S} - \lambda''_{S}) 
&> (\lambda_{S1} - \lambda'_{S1})^{2},
\qquad & 4 \lambda_{2} (\lambda_{S} + \lambda'_{S} - \lambda''_{S}) 
  &> (\lambda_{S2} - \lambda'_{S2})^{2}.
\end{aligned}
\label{eq:bound:cond:CP:inv:Z2:V}
\end{equation}
Because there are more scalar couplings than in the SM, they can counteract the top quark Yukawa term in the beta 
function for $\lambda_{1}$ presented in the next Section and 
lower the vacuum stability bound on $M_{H}$ below the LEP2 experimental bound of 114.4~GeV 
consistently with \Eq{eq:bound:cond:CP:inv:Z2:V}. Therefore, in our model,  the precision data indications for 
light SM Higgs boson does not contradict with vacuum stability constraints.

\section{RGE analyses}
Because our low energy model is induced by GUT, it must stay perturbative up to the GUT scale. 
This requirement implies stringent constraints on the model parameters and consequently on the properties of DM. 
We have derived \cite{RGE} the full set of RGEs of the model \Eq{V}. The one-loop 
beta functions are given by
\begin{equation}
\begin{split}
  \beta_{\lambda_{1}} &= 24 \lambda_{1}^{2} + 2 \lambda_{3}^{2} + 2 \lambda_{3} \lambda_{4} + \lambda_{4}^{2} + \lambda_{5}^{2} + \lambda_{S1}^{2} + \lambda_{S1}^{\prime 2} \\
 &+ \frac{3}{8}(3g^4 + g^{\prime 4} +2g^2 g^{\prime 2}) - 3\lambda_1 (3g^2 +g^{\prime
2}-4y_{t}^2)-6 y_{t}^4,
  \\
  \beta_{\lambda_{2}} &= 24 \lambda_{2}^{2} + 2 \lambda_{3}^{2} + 2 \lambda_{3} \lambda_{4} + \lambda_{4}^{2} + \lambda_{5}^{2} + 
\lambda_{S2}^{2} + \lambda_{S2}^{\prime 2} \\
&+ \frac{3}{8}(3g^4 + g^{\prime 4} +2g^2 g^{\prime 2}) -3\lambda_2
(3g^2 +g^{\prime 2}),
\\
  \beta_{\lambda_{3}} &= 4 (\lambda_{1} + \lambda_{2}) (3\lambda_{3} +  \lambda_{4}) + 4 \lambda_{3}^{2} + 2 \lambda_{4}^{2} + 2 \
\lambda_{5}^{2} + 2 \lambda_{S1} \lambda_{S2} + 2 \lambda'_{S1}
\lambda'_{S2} \\
&+ \frac{3}{4}(3g^4 + g^{\prime 4} -2g^2 g^{\prime 2}) - 3\lambda_3
(3g^2 +g^{\prime 2}-2y_{t}^2),
  \\
  \beta_{\lambda_{4}} &= 4(\lambda_{1} + \lambda_{2}) \lambda_{4} + 8 \lambda_{3} \lambda_{4} + 4 \lambda_{4}^{2} + 8 \lambda_{5}^{2} 
  \\
  & + 3g^2 g^{\prime 2} - 3\lambda_4 (3g^2 +g^{\prime
2}-2y_{t}^2),
  \\
  \beta_{\lambda_{5}} &= 4 (\lambda_{1} + \lambda_{2} + 2 \lambda_{3} 
  + 3 \lambda_{4}) \lambda_{5} \\
  &- 3\lambda_5 (3g^2 +g^{\prime 2}-2y_{t}^2),
\\
  \beta_{\lambda_{S}} &= 20 \lambda_{S}^{2} + 2 \lambda_{S1}^{2} + 
  \lambda_{S1}^{\prime 2} + 2 \lambda_{S2}^{2} + \lambda_{S2}^{\prime 2} 
  + 36 \lambda_{S}^{\prime 2} + \frac{27}{2} \lambda_{S}^{\prime\prime 2} ,
  \\
  \beta_{\lambda'_{S}} &= \lambda_{S1}^{\prime 2} + \lambda_{S2}^{\prime 2} + 24 \lambda_{S} \lambda'_{S} + \frac{9}{2} \lambda_{S}^{\prime\prime 2} ,
  \\
  \beta_{\lambda''_{S}} &= 4 \lambda_{S1} \lambda'_{S1} + 4 \lambda_{S2} \lambda'_{S2} + 36 (\lambda_{S} + \lambda'_{S}) \lambda''_{S} ,
  \\
  \beta_{\lambda_{S1}} &= 4 (3 \lambda_{1} + 2 \lambda_{S}  + \lambda_{S1}) \lambda_{S1} + 4 \lambda_{S1}^{\prime 2} + (4 \lambda_{3} + 2  \lambda_{4}) \lambda_{S2} + 6 \lambda'_{S1} \lambda''_{S} \\
  & - \frac{3}{2} (3 g^{2} + g^{\prime 2} - 4 y_{t}^{2}) \lambda_{S1},
  \\
  \beta_{\lambda_{S2}} &= 4 (3 \lambda_{2} + 2 \lambda_{S} + \lambda_{S2}) \lambda_{S2}  + 4 \lambda_{S2}^{\prime 2} +  (4 \lambda_{3} + 2 \lambda_{4}) \lambda_{S1} + 6 \lambda'_{S2} \lambda''_{S} \\
 & - \frac{3}{2} (3 g^{2} + g^{\prime 2}) \lambda_{S2},
\\
  \beta_{\lambda'_{S1}} &= (4 \lambda_{3} + 2 \lambda_{4}) \lambda'_{S2} + 4 (3 \lambda_{1} + \lambda_{S} + 2 \lambda_{S1} + 3 \lambda'_{S}) \lambda'_{S1} + 6 \lambda_{S1} \lambda''_{S} \\
  & - \frac{3}{2} (3 g^{2} + g^{\prime 2} - 4 y_{t}^{2}) \lambda'_{S1},
  \\
  \beta_{\lambda'_{S2}} &= (4 \lambda_{3} + 2 \lambda_{4}) \lambda'_{S1} + 4 (3 \lambda_{2} + \lambda_{S} + 2 \lambda_{S2} + 3 
\lambda'_{S})  \lambda'_{S2} + 6 \lambda_{S2} \lambda''_{S} \\
 & - \frac{3}{2} (3 g^{2} + g^{\prime 2}) \lambda'_{S2}.  \\
  \beta_{\mu_{1}^{2}} &= 12 \mu_{1}^{2} \lambda_{1} 
  + 4 \mu_{2}^{2} \lambda_{3} + 2 \mu_{2}^{2} \lambda_{4} + 2 \mu_{S}^{2} \lambda_{S1} + 2 \mu_{S}^{\prime 2} \lambda'_{S1} + \frac{1}{2} \left(\mu_{SH}^{2} + \mu_{SH}^{\prime 2} \right) \\
 & -\frac{3}{2} \mu_{1}^{2} (3 g^{2} + g^{\prime 2} - 4 y_{t}^{2}),
  \\
  \beta_{\mu_{2}^{2}} &=  12 
  \mu_{2}^{2} \lambda_{2} + 4 \mu_{1}^{2} \lambda_{3} + 2 \mu_{1}^{2} \lambda_{4} + 2
\mu_{S}^{2} \lambda_{S2} + 2 \mu_{S}^{\prime 2} \lambda'_{S2} + \frac{1}{2} \left(\mu_{SH}^{2} + \mu_{SH}^{\prime 2}\right) \\
& -\frac{3}{2} \mu_{2}^{2} (3 g^{2} + g^{\prime 2}),
\\
  \beta_{\mu_{S}^{2}} &=  8 \mu_{S}^{2} \lambda_{S} + 4 \mu_{1}^{2} \lambda_{S1} + 4 \mu_{2}^{2} \lambda_{S2} 
  + 6 \mu_{S}^{\prime 2} \lambda''_{S} + \mu_{SH}^{2} + \mu_{SH}^{\prime 2},
  \\
  \beta_{\mu_{S}^{'2}} &= 4 \mu_{S}^{\prime 2}\lambda_{S} 
  + 4 \mu_{1}^{2} \lambda'_{S1} + 4 \mu_{2}^{2} \lambda'_{S2} 
  + 12 \mu_{S}^{\prime 2} \lambda'_{S} + 6 \mu_{S}^{2} \lambda''_{S} + 2 \mu_{SH} \mu'_{SH},
  \\
  \beta_{\mu_{SH}} &= 2 \mu_{SH} (\lambda_{3} + 2 \lambda_{4} +\lambda_{S1} + \lambda_{S2}) + 2 \mu'_{SH} (3 \lambda_{5} 
  + \lambda'_{S1} + \lambda'_{S2}) \\
  &-\frac{3}{2} \mu_{SH} (3 g^2 + g^{\prime 2} - 2 y_{t}^2),
  \\
  \beta_{\mu'_{SH}} &= 2 \mu'_{SH} (\lambda_{3} + 2 \lambda_{4} +\lambda_{S1} + \lambda_{S2}) + 2 \mu_{SH} (3 \lambda_{5} 
  + \lambda'_{S1} + \lambda'_{S2}) \\
  &-\frac{3}{2} \mu'_{SH} (3 g^2 + g^{\prime 2} - 2 y_{t}^2).
\label{eq:betas:masses}
\end{split}
\end{equation}
We also include the one-loop $\beta$-functions for $g$, $g'$, $g_{3}$ and $y_{t}$, given by
\begin{equation}
\begin{split}
  \beta_{g'} &= 7 g^{\prime 3} ,\\
  \beta_{g} &= -3 g^{3}, \\
  \beta_{g_{3}} &= -7 g^{3}, \\
  \beta_{y_{t}} &= y_{t} \left( \frac{9}{2} y_{t}^{2} - \frac{17}{12} g^{\prime 2} - \frac{9}{4} g^{2} - 8 g_{3}^{2} \right).
\end{split}
\end{equation}
Based solely on the running due to the low-energy RGEs, 
we identify the unification scale $2\times 10^{16}$~GeV by the RGE solution for $g_2=g_3.$
The exact values of gauge couplings at $M_G$ are given by $g_1= \sqrt{5/3} g' = 0.58,$ $g_2=g_3=0.53.$
Based solely on the running due to the low-energy RGEs, 
the precision of unification of all three gauge couplings in our model is better than in the SM because of the existence of an
extra scalar doublet.  We assume that an exact unification can be achieved due to the GUT thresholds corrections in full 
$SO(10)$ theory which we cannot estimate because the details of GUT symmetry breaking are not known.
Those corrections can have only small logarithmic influence on our numerical estimates of $g_i$ just below $M_G$ and
affect our numerical results for DM negligibly.
In our numerical analysis we follow the strategy used in similar studies of parameter running in SUSY GUT theories.
We fix all the measured model parameters and the SM Higgs boson mass at $M_Z$ and calculate 
the corresponding $\mu_1^2$ and $\lambda_1.$ We run them up to the GUT scale where we 
randomly generate new physics parameters assuming the $SO(10)$ boundary conditions. 
We iterate running until the relative error between fixed and calculated $\mu_1^2$ and $\lambda_1$ at $M_{Z}$ gets smaller than 1\%.
After that we calculate the DM abundance and direct detection
cross section at $M_{Z}$.

We find that the new physics parameters are strongly constrained by 
 the vacuum stability and perturbativity arguments. For example, assuming all $\lambda_i$ allowed by \Eq{VGUT} to be
equal at $M_Z,$ perturbativity of them up to $M_G$ requires $\lambda_i (M_Z)< 0.194, \, 0.187,\, 0.170$
for $M_H=120,\, 140,\, 160$~GeV, respectively.
We also impose the GUT boundary conditions~\Eq{bc1}.

\begin{figure}[t]
\includegraphics[width=0.8\textwidth]{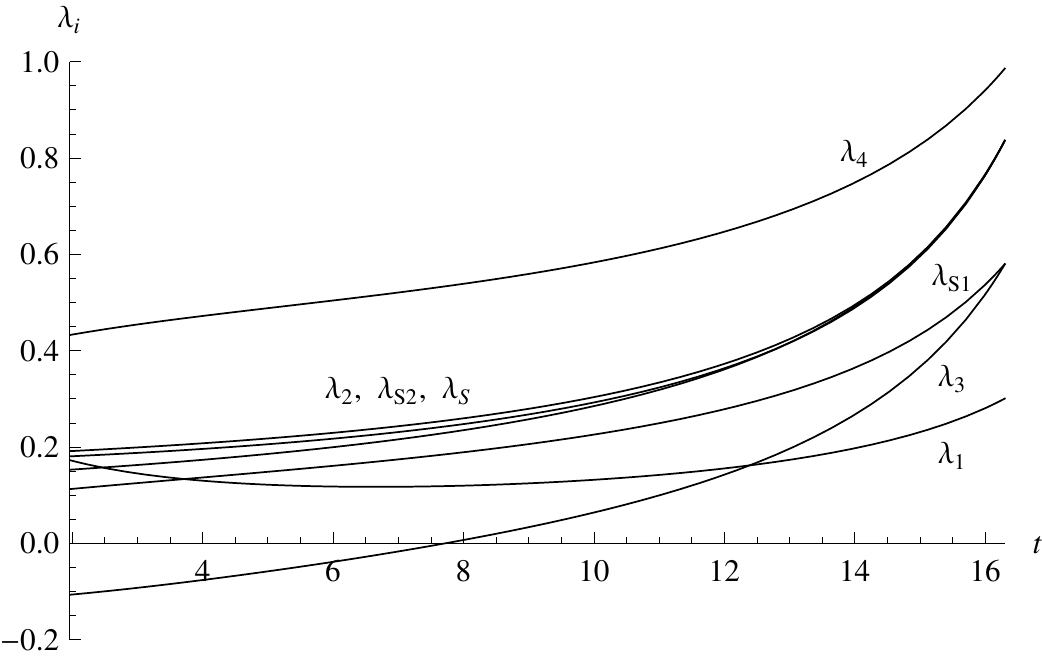}
\caption{An example of $\lambda_i$ running from $M_G$ to $M_Z.$
All $\lambda_{i}$ not suppressed by $SO(10)$ boundary conditions \Eq{bc2} are shown.}
\label{fig1}
\end{figure}

\begin{figure}[t]
\includegraphics[width=0.8\textwidth]{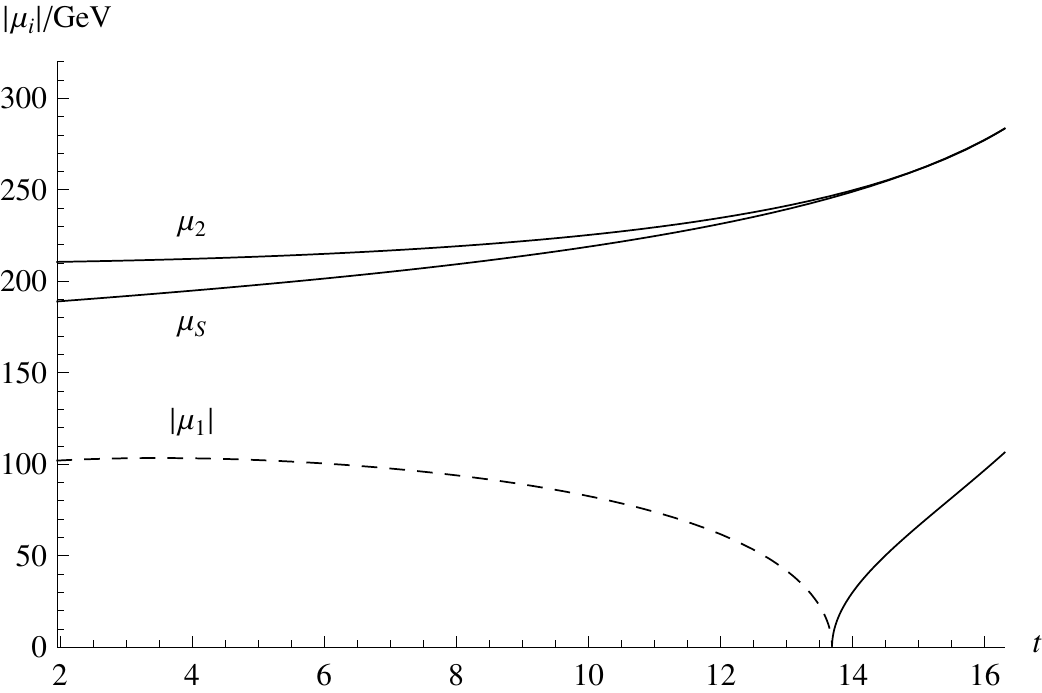}
\caption{ An example of $\mu_{1,2,S}$ running from $M_G$ to $M_Z.$ Dashed line represents negative values of $\mu^2_1$
inducing EWSB.}
\label{fig2}
\end{figure}

We present one consistent example of the evolution of scalar self couplings $\lambda_i$ 
and mass parameters $\mu_i$ between $M_Z$ and $M_G$ in Fig.~\ref{fig1} and in Fig.~\ref{fig2}, respectively. 
We assume $M_H=140$~GeV and take $\mu_{SH}'(M_G)=1$~GeV.
The couplings $\lambda_i(M_Z)$ must be small as not to reach the Landau pole below $M_G.$
The SM gauge symmetry $SU(2)_L\times U(1)_Y$ is not broken at $M_G$ because
all scalar mass parameters are positive $\mu^2_i(M_G)>0.$
However, the SM Higgs mass parameter $\mu_1^2$ exhibits stronger running than the DM mass
 parameters and triggers the radiative EWSB as in SUSY models \cite{Ibanez:1982fr}. 
 In our case the EWSB is 
induced by DM couplings to the SM Higgs boson. (Previously, EWSB via a Coleman-Weinberg-like mechanism has been considered in the Inert Doublet Model \cite{Hambye:2007vf}.)

An interesting feature of the model, demonstrated in Fig.~\ref{fig2}, is that the singlet DM mass parameter
at low energies is always smaller than the doublet one, $\mu^2_S(M_Z)<\mu^2_2(M_Z).$ 
Thus, for small singlet-doublet mixing as is assumed in this example, the DM particle is 
predominantly scalar singlet $S$ whose real and imaginary component mass degeneracy is lifted by
small $ |\mu_S^{\prime 2}|\ll |\mu_1^2|.$

\section{Predictions for DM mass and direct detection cross section}
We assume that DM is a thermal relic and calculate its abundance and direct cross section with matter 
using the MicrOMEGAs package~\cite{micromegas}. The DM interactions \rfn{V} were calculated using the FeynRules 
package~\cite{Christensen:2008py}. We scan over the entire parameter space satisfying \Eq{bc1}, and calculate the RGE evolution of those parameters down to the EW scale.

\begin{figure}[t]
\includegraphics[width=0.8\textwidth]{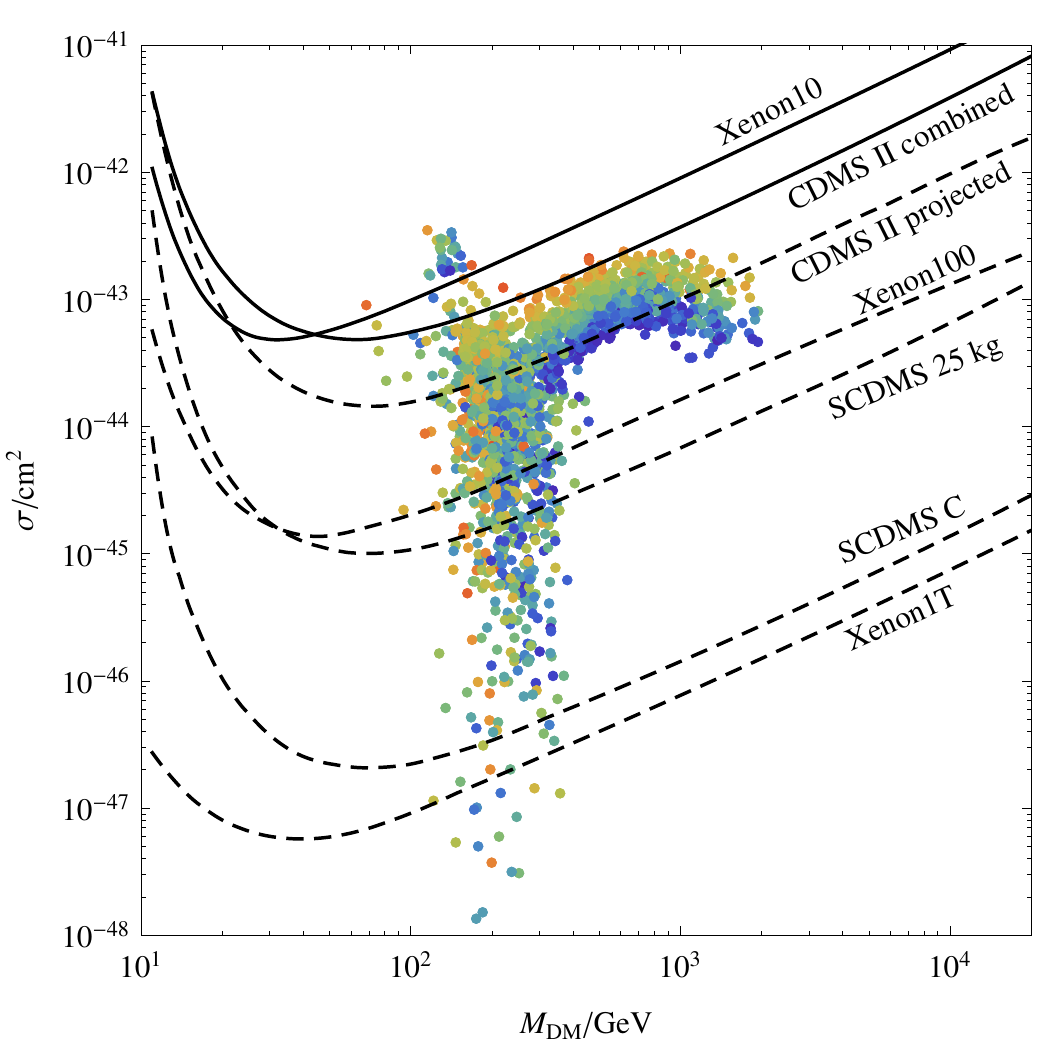}
\caption{ DM direct detection cross-section per nucleon~{\it vs}.~$M_{\mathrm{DM}}$. Color shows SM Higgs masses from 115~GeV (red) to 170~GeV (violet). The points shown encompass the whole parameter space allowed by theoretical and experimental constraints.
}
\label{fig3}
\end{figure}

 Fig.~\ref{fig3} presents a scattered plot of the spin-independent DM 
direct detection cross section per nucleon as a function of DM mass 
$M_{\mathrm{DM}}$ for the SM Higgs boson mass range from 115~GeV (red) to
170~GeV (violet).
The whole parameter space allowed by theoretical constraints of vacuum stability and positive masses and experimental constraints from LEP2 and the WMAP $3\sigma$ result $0.094<\Omega_{DM} h^2 < 0.129$ \cite{Komatsu:2008hk} is shown.
After fixing $\lambda_1$ and $\mu_1^2$ from the assumed SM Higgs boson mass,  we randomly generate the remaining scalar self couplings and mass parameters at $M_G$ in the  ranges $0< |\lambda_i|<4\pi$ and $0< \mu^2_i<(10$~TeV$)^{2}$.
After RGE running the numerical ranges for non-zero parameters at $M_{Z}$ are $0.117 < \lambda_{1} < 0.239$, $0.024 < \lambda_{2} < 0.227$, $-0.424 < \lambda_{3} < 0.247$, $-0.584 < \lambda_{4} < 0.599$, $0.037 < \lambda_{S} < 0.177$, $0.000 < \lambda'_{S} < 0.098$, $-0.212 < \lambda_{S1} < 0.221$, $0.031 < \lambda_{S2} < 0.234$, $-14548~\gev^{2} < \mu_{1}^{2} < -7056~\gev^{2}$, $3147~\gev^{2} < \mu_{2}^{2} < 1.72 \times 10^{7}~\gev^{2}$, $2894~\gev^{2} < \mu_{S}^{2} < 3.84 \times 10^{6}~\gev^{2}$, $-11634~\gev < \mu'_{SH} < 11504~\gev$.
 
 The present \cite{ddnow} and future \cite{ddfut} experimental sensitivities for DM direct detection are  shown in Fig.~\ref{fig3}.
As a result,  
we find that the DM mass is restricted to the window 70~GeV$\lsim M_{\mathrm{DM}} \lsim $2~TeV. The lower bound comes from
 non-observation of charged scalars at LEP2.
 The upper bound $M_{\mathrm{DM}}\lsim 2$~TeV comes from the the requirement of perturbativity of the model parameters 
 up to $M_G.$  Therefore, the DM mass scale $M_{\mathrm{DM}}\lsim {\cal O}(0.1-1)$~TeV is a {\it prediction}
 of our scalar DM GUT scenario.

The direct DM interaction with nuclei occurs via the SM Higgs boson exchange. 
The dominant DM-Higgs effective coupling involved in this process is
\begin{equation}
\lambda_{\text{eff}} \, v = \frac{1}{2} (\sqrt{2} s\, c\, \mu'_{SH} + 2 s^{2} (\lambda_{3}+\lambda_{4}) v +2 c^{2} \lambda_{S1} v),
\label{lameff}
\end{equation}
where $s,\,c$ are the sine, cosine of the singlet-doublet mixing angle.
If $M_{\mathrm{DM}}\lsim 300$~GeV, cancellation between different terms in \Eq{lameff} is possible and the spin independent direct
detection cross section can be accidentally small, {\it cf.}  Fig.~\ref{fig3}.
However, for larger DM masses both \Eq{lameff} and thermal freeze-out cross section are
 dominated by large $\mu'_{SH}$ term and one obtains 
a relation between the DM abundance and the direct detection cross sections with only mild dependence on $M_H$
via RGEs. For $M_{\mathrm{DM}}=1$~TeV the WMAP result predicts {\it a lower bound} 
$\sigma/n > 2\cdot 10^{-43} (115~\mrm{GeV}/M_H)^4$~cm$^2$ which is well within the reach of the planned experiments, 
 {\it cf.}  Fig.~\ref{fig3}. Should CDMS experiment observe DM scattering on nuclei, light SM Higgs boson mass is indicated 
 in agreement with precision electroweak data.

\section{DM indirect detection}
The PAMELA~\cite{pam}, ATIC~\cite{atic}, HESS~\cite{Aharonian:2009ah}
 and Fermi~\cite{Abdo:2009zk} anomalies of cosmic ray positron/electron fluxes
can be explained with $ {\cal O}(1)$~TeV mass DM decays~\cite{dmdecay} via $d=6$ operators~\cite{Arvanitaki:2008hq},
preferably to multi-particle final state~\cite{Arvanitaki:2009yb,Meade:2009iu}. 
Non-observation of photons associated with DM annihilation in the Galactic center~\cite{Bertone:2008xr}
and in DM haloes in the Universe~\cite{Huetsi:2009ex} as well as the suppression of hadronic annihilation 
modes~\cite{Cirelli:2008pk} strongly favor DM decays over annihilations as a solution to the anomalies.

In our scenario the decays of DM are most naturally explained via the seesaw like operator
$LLH_1H_2$ which, in addition to the suppression by heavy
Majorana neutrino scale $M_N,$ must be suppressed by the $Z_2$ breaking effects by
additional $M_P.$ We obtain that below EWSB scale the dominant decay mode is given by
\bea
\frac{\lambda_N}{M_N}\frac{m}{M_{P}} LL H_1 H_2 \to 10^{-30}~ \mrm{GeV}^{-1} \nu l^- W^+ H_2^0,
\label{decay}
\eea
where we have taken $\lambda_N\sim 1,$ $M_N\sim 10^{14}$~GeV
and $m\sim v\sim 100$~GeV. In the decays of $W^+$ antiprotons are produced about in 10\%
of decays. Such a small fraction of antiprotons is till allowed by PAMELA data taking 
into account uncertainties in the cosmic ray propagation models \cite{Donato:2008jk}. 
Such a small effective Yukawa coupling of \Eq{decay} can explain the long DM lifetime $10^{26}$~s
without conflicting with the present observational constraints.

\section{Conclusions}
We have argued that the existence of DM, the baryon asymmetry of the Universe, and small neutrino masses
may all signal the same underlying GUT physics. Although $B-L$ is broken in Nature by heavy neutrino Majorana masses, 
$Z_2$ parity $(-1)^{3(B-L)}$ is respected by interactions of all matter fields.
Hence, group theory predicts that in $SO(10)$ GUTs the non-supersymmetric DM must be contained 
in the scalar representation $\bf 16.$   

Based on $SO(10)$ GUT, we have presented a minimal DM model, calculated the full set of its RGEs and studied its
predictions.
 We find that the EWSB occurs radiatively due to SM Higgs boson couplings to the DM, analogously to SUSY models.
 The thermal relic DM mass is predicted to be $M_{\mathrm{DM}}\lsim {\cal O}(0.1-1)$~TeV by the requirement of
 perturbativity of model parameters up to the GUT scale. If $M_{\mathrm{DM}}\gsim 300$~GeV as suggested by DM decay
 solution to the recently observed cosmic ray anomalies, the WMAP measurement of DM abundance
 predicts a lower bound on DM spin independent direct cross section with nuclei,
 which is within the reach of planned experiments for all values of the SM Higgs boson mass.

\vskip 0.3 cm

\noindent {\bf Acknowledgment.}
We thank M. Tytgat for communication. This work was supported by the ESF Grant 8090 and
by EU FP7-INFRA-2007-1.2.3 contract No 223807.


\end{document}